\documentclass[twocolumn]{aastex63}

\usepackage{amsfonts,amsmath,graphicx,natbib,subfigure,color,gensymb,longtable,footnote,soul,lineno}

\received{}
\revised{ -- }
\accepted{}
\submitjournal{ApJ}

\shorttitle{Photometric redshift estimation}
\shortauthors{Baldeschi et al. }

\graphicspath{{./}{figures/}}

\begin{document}

\title{Photometric redshift estimation of galaxies in the P\lowercase{an}-STARRS 3$\pi$ survey- I. Methodology}

\correspondingauthor{Adriano Baldeschi}
\email{adriano.baldeschi@northwestern.edu}

\author{Baldeschi A.}
\affiliation{Center for Interdisciplinary Exploration and Research in Astrophysics (CIERA) and Department of Physics and Astronomy, Northwestern University, Evanston, IL 60208}

\author{Stroh M.}
\affiliation{Center for Interdisciplinary Exploration and Research in Astrophysics (CIERA) and Department of Physics and Astronomy, Northwestern University, Evanston, IL 60208}

\author[0000-0003-4768-7586]{Margutti R.}
\altaffiliation{Alfred P. Sloan Fellow.}
\affiliation{Center for Interdisciplinary Exploration and Research in Astrophysics (CIERA) and Department of Physics and Astronomy, Northwestern University, Evanston, IL 60208}
\affiliation{CIFAR Azrieli Global Scholars program, CIFAR, Toronto, Canada}

\author{Laskar, T.}
\affiliation{Department of Physics, University of Bath, Claverton Down, Bath, BA2 7AY, UK}

\author[0000-0001-9515-478X]{A.~A.~Miller}
\affiliation{Center for Interdisciplinary Exploration and Research in Astrophysics (CIERA) and Department of Physics and Astronomy, Northwestern University, Evanston, IL 60208}
\affiliation{The Adler Planetarium, Chicago, IL 60605, USA }




\begin{abstract}
We present a  photometric redshift (photo-$z$) estimation  technique  for galaxies in the P\lowercase{an}-STARRS1 (PS1) $3\pi $ survey. Specifically,  we train and test a regression and a classification  Random-Forest (RF)  models  using photometric features (magnitudes, colors and moments of the radiation intensity) from the optical PS1 data release 2 (PS1-DR2) and from the AllWISE/unWISE infrared source catalogs. The classification RF model ($RF_{clas}$) has better performance in the local universe ($z\lesssim 0.1$), while the second one ($RF_{reg}$)
is on average better for $0.1 \lesssim z\lesssim1$. 
We adopt as labels the spectroscopic redshift of the galaxies from the Sloan Digital Sky Survey (SDSS) data release 16 (SDSS-DR16). We find that the combination of AllWISE/unWISE  and PS1-DR2 features leads to an average bias  of $\overline{\Delta z_{norm}}=1\times 10^{-3}$, a standard deviation $\sigma(\Delta z_{norm})=0.0225$, (where $\Delta z_{norm}  \equiv  (z_{phot}-z_{spec})/(1+z_{spec})$), and an outlier rate of  $P_0=1.48 \%$ in the test set for the $RF_{clas}$ model.  In the low-redshift Universe ($z<0.1$) that is of primary interest to many astronomical transient studies, our model produces an error estimate on the inferred magnitude of an object of $\le$1 mag  in 87\%  of the test sample.
\end{abstract}


\section{Introduction} \label{sec:intro}
The photometric redshift (photo-$z$) provides estimates of the distance of astronomical objects, and is an essential tool in multiple areas of astronomy. The main benefit of photo-$z$ is that distance estimates are obtained rather inexpensively for the sources detected in the images as opposed to spectroscopic redshift determination. The drawback is that  photo-$z$ estimates typically have  lower accuracy when compared with their spectroscopic counterparts. Photo-$z$ estimation is critical for a number of endeavours, including astronomical transient typing, constraining the presence dark energy (DE) with weak-lensing, and can also be employed for other DE probes like supernovae (SNe)  type Ia, the mass function of galaxy clusters and galaxy clustering as well (e.g., \citealt{Salvato2019}). 
In this work we  focus  on photo-$z$ for transient classification
but we note this methodology can be extended to other areas of astronomy.
Host galaxy redshift,  as highlighted by \citet{Muthukrishna+2019}, is  one of the most predictive contextual-information features for  transient classification, because it enables a direct estimate of the intrinsic luminosity of the transient.

Astronomical transients  are historically  classified according to their spectral and photometric features. The wide field of view, high cadence and sensitivity of astronomical surveys is currently leading to the discovery of  thousands of new transient events every night.  This plethora of new transient discoveries is currently made possible thanks to surveys such the Zwicky Transient Facility (ZTF, \citealt{Bellm19}), the Young Supernova Experiment (YSE, \citealt{Jones2020}) and
  the  Asteroid Terrestrial-impact Last Alert System (ATLAS, \citealt{Tonry2018}). 
Forthcoming  surveys such as the Legacy Survey of Space and Time (LSST, \citealt{Ivezic2019}) carried out on   Vera C. Rubin Observatory will further  enhance  the rate of discovery  of new transients,  
which will make prompt spectroscopic classification of the majority of transients  unfeasible. It is thus crucial  to establish  new paths for  transient classification. 

The  two alternative  classification methods consist of leveraging the transient photometry \citep[e.g.][]{Newling2011,Karpenka2013,Moller2016,Lochner+2016,Sooknunan2018,Narayan2018,Pasquet2019,Muthukrishna+2019,Ishida2019,Villar+2019,Moller2020} 
and the contextual information of the environments (e.g., redshift, host-galaxy morphology and star formation rate) where the transients happen
\citep{Foley+2013,Baldeschi2020,Gagliano2020}  by using 
machine learning (ML) algorithms. In this work we focus on the estimation of a primary contextual parameter, the host-galaxy redshift, using available optical and infra-red band photometry.

The methods used to infer galactic photometric redshifts fall into two broad categories: (i) methods based  on physical modeling of the multi-band emission from galaxies, and (ii) supervised  ML-based methods. An  extensive recent review of the literature for both methods can be found in \citet{Salvato2019}. Physical model-based algorithms \citep[e.g.][]{Benitez2000,Arnouts2002,Ilbert2006,Beck2016} rely on the template-fitting approach where observed photometric data are compared to simulated photometry for a wide number of template galaxy spectra and redshifts.
Supervised ML methods \citep[e.g.,][]{Pasquet2019,Zhou2020,Tarrio2020,Ansari2020,Schuldt2020}, on the other hand, require large training sets of spectroscopic redshifts that are used to infer an  intrinsic correlation with the photometric features (e.g., colors, magnitudes and shape parameters like moments of radiation intensity) in a data-driven fashion. Both physical model-based and ML-based algorithms
are not yet competitive with the accuracy and precision of spectroscopic redshifts across the entire range of $z$ where galaxies are detected and known to exist (i.e. in the local and in the distant Universe). 

There are three main issues with the photo-$z$ estimation using ML-based techniques: (i) the redshift of galaxies in the local Universe ($z\lesssim 0.1$) are typically overestimated (i.e. the inferred photo-$z\gtrsim 0.1$);
(ii) the redshift of galaxies at larger $z$ ($z\gtrsim 0.7$) are typically underestimated (i.e. the inferred photo-$z>0.7$);
(iii) presence of catastrophic outliers (i.e. sources for which the photo-$z$ estimate significantly differ from the true spectroscopic redshift).
In this work we attempt to mitigate the effects of the three issues above, with special focus on the use of photo-$z$ by the astronomical transient community in the local Universe.  Specifically, we train and test two Random-Forest (RF) models leveraging the PS1-DR2, AllWISE and the unWISE photometric features (magnitudes, colors and moments of the radiation intensity) and leveraging the SDSS labels (spectroscopic redshift of the galaxies). The combination of AllWISE, unWISE and PS1-DR2 photometry that covers the optical and infrared (IR)  leads to 
improvements when compared to other studies that used the PS1-DR2 photometry
alone \citep[e.g.,][]{Pasquet2019,Tarrio2020,Beck2020}.

This work is the first of a series of two papers where we 
focus on the development of the ML model. In a forthcoming study, we will present
a catalog with the photo-$z$ estimate of PS1-DR2 galaxies.
The paper is organized as follows. In \S\ref{datdesc} we describe the four datasets used (SDSS-DR16, PS1-DR2, AllWISE, unWISE).
In \S\ref{ML} we create a training/testing  set, we pre-process the data and we develop two RF models.  In \S\ref{mainqq} we discuss  our main  results.
Conclusions are drawn in \S\ref{conclusions}.

\section{Datasets description}
\label{datdesc}
In our analysis we use four catalogs: (i) the data release 16 of the the Sloan Digital Sky Survey (SDSS-DR16, \citealt{Ahumada2020}); (ii) the second P\lowercase{an}-STARRS1  (PS1, \citealt{Chambers+2016}) data release  of the $3\pi $ survey (PS1-DR2 hereafter); (iii) the AllWISE source catalog \citep{Wright2010};  (iv) the unWISE source catalog \citep{Schlafly2019}.

 PS1 data have been collected with a 1.8 meter telescope to produce images of the sky through five filters (center wavelengths: $y_{P1}$ [9633 {\AA}], $z_{P1}$ [8679 {\AA}], $i_{P1}$ [7545 {\AA}], $r_{P1}$ [6215 {\AA}], $g_{P1}$ [4866 {\AA}]). Two surveys have been completed with  PS1: the 3$\pi$ survey (3$\pi$S) and the medium deep survey. Here, we utilize data from the 3$\pi$S, which covers the sky northern  of declination $\delta=  - 30^\circ$ and includes  data collected between 2009-06-02 and 2014-03-31. The limiting magnitudes of the 3$\pi$S are $\sim$21.5 and $\sim$22.5  mag for the  $y_{P1}$ and $z_{P1}$, respectively, while it is $\sim$23.5 mag for the  $i_{P1}$, $r_{P1}$ and $g_{P1}$  filters. 
In this paper, we use the PS1-DR2 data from the  ``StackObjectAttributes'' table\footnote{\href{https://outerspace.stsci.edu/display/PANSTARRS/}{StackObjectAttributes table link}}
that contains  photometric information (e.g., PSF-flux, Kron-flux) of the stacked  data, estimated as presented in \cite{Magnier2013}.  Sources  included in this table have been detected with a signal-to-noise, $S/N>20$ for each individual exposure. The table also contains   detections of the same source from consecutive exposures, which implies that there can be different  photometric measurement in the same band for a given source. In \S\ref{main}  we use the PS1-DR2 data to
 train our RF model. A detailed description of the meaningful features of this data set is provided in \S\ref{pre_pro}.

The Wide-Field Infrared Survey Explorer (WISE) mapped the sky at $\lambda=$22, 12, 4.6 and 3.4 $\mu m$ (W4, W3, W2, and W1 bands) \citep{Wright2010},  with an angular resolution of 12.0$\arcsec$, 6.5$\arcsec$, 6.4$\arcsec$ and  6.1$\arcsec$ in the four bands, respectively.  AllWISE includes the data acquired during the WISE full cryogenic mission phase, which was carried out between 7 January 2010 and 6 August 2010. The AllWISE  data release consists of  coadded and calibrated  images and a catalog with  photometric and positional  information for $\approx 563$ million sources found in the WISE images \citep{Wright2010}. 
The unWISE catalog includes the  fluxes of  two billion objects
observed by  WISE over the entire sky \citep{Schlafly2019}.
The unWISE catalog has two  advantages over  AllWISE: (i) it is based on  deeper
imaging; (ii) it features a better modeling of crowded regions of the sky.
However, a clear disadvantage of unWISE is that the fluxes are available  at 3.4 and  4.6 $\mu m$, only.

The fourth catalog that we use is derived from the Sloan Digital Sky Survey (SDSS), which  has been observing from the Apache Point Observatory (APO) since 1998 using a 2.5 m  telescope \citep{Gunn2006} and from Las-Campanas Observatory (LCO) since 2017 using a 2.5 m telescope. SDSS  produces images of the sky through five  filters
($z_{S16}$ [9134 {\AA}], $i_{S16}$ [7625 {\AA}], $r_{S16}$ [6231 {\AA}], $g_{S16}$ [4770 {\AA}] ,$u_{S16}$ [3543 {\AA}]). The SDSS-DR16 catalog  \citep{Ahumada2020} provides the spectroscopic  redshift of the galaxies with  $z\lesssim  1$ that we  use as labels throughout the paper.

\begin{figure}
\centering
\includegraphics[width=0.45\textwidth]{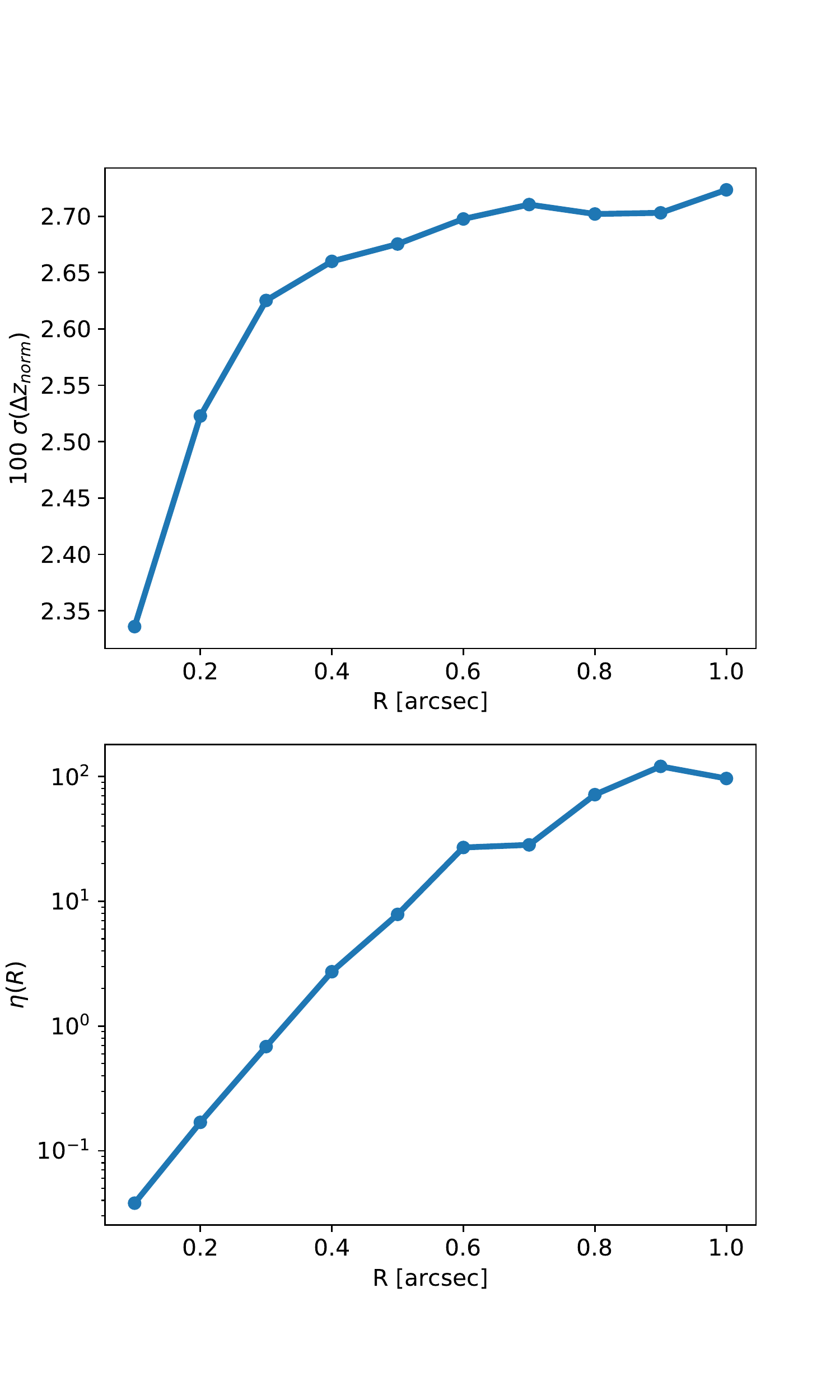}
\caption{ \emph{Upper panel}: $\sigma (\Delta z_{norm})$ vs the crossmatching radius
between the PS1-DR2  and the SDSS-DR16 sources. The $\sigma (\Delta z_{norm})$ has been estimated in  tests set  by leveraging a $RF_{reg}$ model with 10 trees for each crossmatching radius.
\emph{Lower panel}: The contamination, $\eta(R) =\frac{\Delta N_{random}(R)}{\Delta N_{catalog}(R)}$, as a function of the search radius $R$ between the PS1-DR2 and the SDSS-DR16 catalogs.
}
\label{fig:sdss_ps1dr2_xmatch}
\end{figure}

\begin{figure}
\centering
\includegraphics[width=0.5\textwidth]{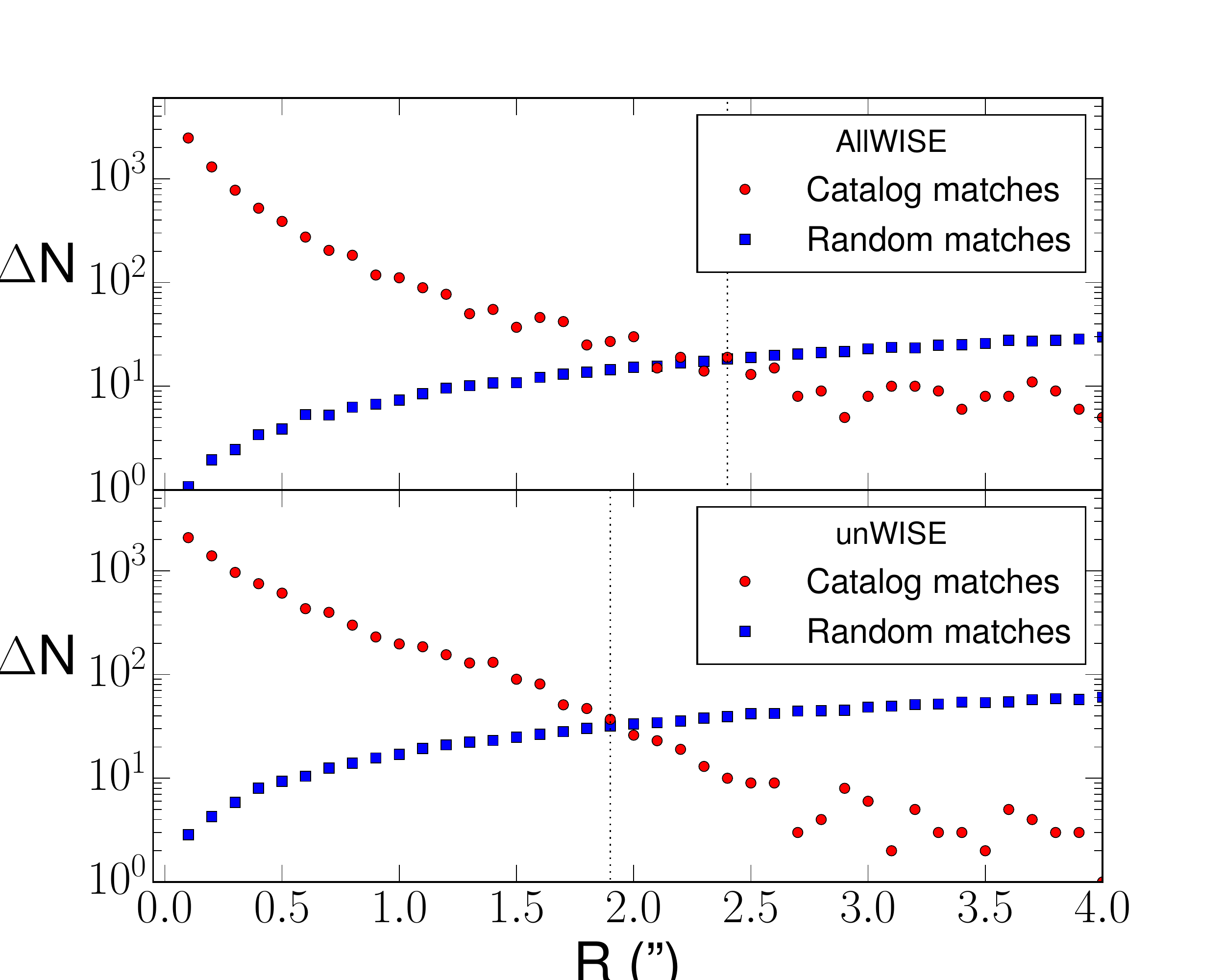}
\caption{\textit{Top panel}: $\Delta N$, the number of additional cross-matches, between PS1-DR2 and AllWISE at $0.1''$ search radii  intervals. 
The vertical dashed line at $R=2.4''$ highlights the radius at which the average number of random cross matches to AllWISE is greater than the number of matches between PS1-DR2 and AllWISE.
\textit{Bottom panel}: $\Delta N$ for cross matches between PS1-DR2 and AllWISE at different search radii at 0.1'' intervals.
The vertical dashed line at $R=1.9''$ highlights the radius at which the average number of random cross matches to unWISE is greater than the number of matches between PS1-DR2 and unWISE.
}
\label{fig:allwise_unwise_crossmatching}
\end{figure}

\section{A machine learning approach to photo-$z$ estimation}
\label{ML}
In this section, we analyze the  PS1-DR2, AllWISE, unWISE  and SDSS-DR16 catalogs. We  cross-match the  catalogs and for each common source we retain features that are meaningful to our subsequent analysis. We then introduce the RF classification and regression  models for photo-$z$ estimation. Finally, we present the metrics to rank the model.

\subsection{Training Sample Selection}
\label{sec:sample_selection}
We analyzed the accuracy of the cross matching between SDSS-DR16 and PS1-DR2 by comparing the number of positional cross matches to PS1-DR2 sources using SDSS-DR16 positions and random angular offsets similar to approaches used by \citet{Stephen2006}, \citet{DAbrusco2013}, and \citet{DAbrusco2014}.
The SDSS-DR16 dataset contains both galaxies and quasi-stellar object (QSO), which are labeled as \texttt{GALAXY} and \texttt{QSO}, respectively, in the \texttt{CLASS} field of the SDSS-DR16 \texttt{specObj} table. 
We selected $10^4$ random SDSS-DR16 galaxies and QSOs, and we created a random source list consisting of 100 offset positions for each source in the catalog positions resulting in $10^6$ random positions.
Each random offset was chosen to be between $1'$ and $2'$ from the SDSS-DR16 position, and at a random position angle.
The large offsets were chosen so that genuine cross matches would not contribute to the randomly generated cross matching.
Random offsets and position angles were selected from separate uniform random distributions.

We calculated the cumulative number of cross matches between the SDSS-DR16 catalog position and PS1-DR2, $N_{catalog}$, at angular separations, $R$, at intervals $\Delta R=0.1''$. We consider a PS1-DR2 source associated with a SDSS-DR16 source if it is within $R$ of the SDSS-DR16 source. We then calculate the differential number of cross matches at a given angular separation $\Delta N_{catalog}(R)\equiv N_{catalog}(R)-N_{catalog}(R-\Delta R)$.
Similarly, we calculated the differential number of cross matches for each random offset.
Since we calculated 100 random offsets per SDSS-DR16 source, we normalized the number of randomized cross matches to the same sample size as $\Delta N_{catalog}(R)$ by dividing by 100.
Thus, we defined $\Delta N_{random}(R)$ as the average number of cross matches per $10^4$ positions.
In the case of multiple cross matches between the two catalogs, we chose the pair with the closest angular separation.
For $\Delta N_{random}(R)$, we only considered the radius at which the closest association was made, and thus multiple possible associations were only counted at the first possible association.
The evolution of the contamination parameter, defined as $\eta(R) \equiv \frac{\Delta N_{random}(R)}{\Delta N_{catalog}(R)}$,  as a function of the search radius $R$ is shown in Figure \ref{fig:sdss_ps1dr2_xmatch}. As expected the level of contamination increases with larger search radii. 
We ran the ML algorithm defined below using SDSS-DR16 and PS1-DR2 cross match radii from $R=0.1''$ to $R=1.0''$, and found that $\sigma(\Delta z_{norm})$ significantly increases when $\eta\ge0.1\%$, corresponding to SDSS-DR16 and PS1-DR2 cross matching radius  $R\approx 0.1''$ (Figure \ref{fig:sdss_ps1dr2_xmatch}). 
The $\sigma(\Delta z_{norm})$ value is a metric
to evaluate the performance of the model and it will be defined in 
\S \ref{ML2}.
Thus, we adopted an SDSS-DR16 and PS1-DR2 cross matching angular separation of $0.1''$, resulting in a sample size of 3,487,203 SDSS-DR16 sources with a counterpart in PS1-DR2. 

The optimum cross matching angular separations between PS1-DR2 and AllWISE and unWISE catalogs were also examined similar to the SDSS-DR16 and PS1-DR2 method described above.
In this case, we selected $10^4$ random sources from the complete SDSS-DR16 and PS1-DR2 cross matched sample. The PS1-DR2 coordinates were cross matched against AllWISE and unWISE utilizing the identical $10^{4}$ random sources.
The PS1-DR2 positions were chosen so that the same selection could be adopted to the PS1-DR2 sample outside of the SDSS-DR16 footprint.
We adopted the greatest radii at which the number of catalog matches is greater than the number of random cross matches, which corresponds to $R=2.4''$ and $R=1.9''$ for the PS1-DR2 to AllWISE and unWISE cross-matching radii, respectively, (Figure \ref{fig:allwise_unwise_crossmatching}).  

After retaining only those sources labeled as \texttt{GALAXY} in the \texttt{specObj} SDSS-DR16 table, the  data set consists of 2,394,092 sources. In the PS1-DR2 catalog there are often multiple detections of the same source, which create multiple photometry sets for a given source. In case of multiple photometry sets for a given source, we retain the photometry associated with ``primaryDetection'' equal to 1 .
The final SDSS-DR16/PS1-DR2/AllWISE/unWISE data set contains 1,251,249  unique galaxies.

\subsection{Feature Selection and Pre-processing}
\label{pre_pro}

For each  galaxy, we identify the  properties  listed in   PS1-DR2, AllWISE and unWISE that are  meaningful to train a ML model, and associate the labels (redshift) from the SDSS-DR16 catalog.
 For each PS1  photometry filter, we identify the relevant features as  the PSF-mag,  the Kron-mag \citep{Kron+1980}, and the second moment of the radiation-intensity, defined as $<XY>=\int_{Sxy} u v I(u,v) \,du\,dv$ or $<X^2>=\int_{Sx} u^2 I(u) \, du$, where $I$ is the radiation-intensity. The moments of the radiation intensity  are  correlated with the distribution of light in a galaxy and  provide information on the galaxy shape. 
These features   are presented in the ``StackObjectAttributes'' table of  PS1-DR2  as
\texttt{\{g,r,i,z,y\}PSFMag, \{g,r,i,z,y\}KronMag, momentYY, momentXY and momentXX}.

Furthermore, we use the AllWISE photometry: 
\texttt{w1mag, w2mag, w3mag, w4mag, w1mag\_1, w2mag\_1,}
\texttt{w3mag\_1, w4mag\_1, w1mag\_2, w2mag\_2, w3mag\_2, w4mag\_2,
       w1mag\_3, w2mag\_3, w3mag\_3, w4mag\_3,
       w1mag\_4, w2mag\_4,}
\texttt{w3mag\_4, w4mag\_4} as meaningful features. 
The \texttt{w1mag, w2mag, w3mag w4mag} features are the W1, W2, W3 and W4 magnitudes, respectively
while the  \texttt{w\{1,2,3,4\}mag\_\{1,2,3,4\}} are the photometry taken with 
different  extraction apertures.\footnote{See \href{https://wise2.ipac.caltech.edu/docs/release/allwise/expsup/sec2_1a.html}{description} for a comprehensive  description of the WISE photometric catalog.}
We also use the unWISE photometry:
\texttt{unwise\_w1\_mag\_ab} and  \texttt{unwise\_w2\_mag\_ab} that are the $W1_{un}$ and $W2_{un}$ unWISE magnitudes.
The observed magnitudes  are distance-dependent quantities 
and therefore  meaningful features for our ML model.

In addition to the observed mags, the observed colors of galaxies are important tracers of their distance in the universe (through their redshift dependence). For this reason, we add the colors as additional features in our ML model.
 We adopt the following set of  colors as features:
 $g_{P1}-r_{P1}$, $r_{P1}-i_{P1}$, $i_{P1}-z_{P1}$, $z_{P1}-y_{P1}$ for both the Kron and PSF magnitudes. 
  We also create the AllWISE colors as: e.g., W1-W2, W2-W3, W3-W4. 
 We also add a mixture set of AllWISE/PS1-DR2 colors: e.g., $g_{P1}-W1$, $r_{P1}-W1$,  $i_{P1}-W2$, $z_{P1}-W3$. 
 Furthermore, we build the unWISE color $W2_{un}-W1_{un}$ and the unWISE/PS1-DR2 mixture colors:
 e.g., $g_{P1}-W1_{un}$, $r_{P1}-W1_{un}$,  $i_{P1}-W2_{un}$, $z_{P1}-W3_{un}$. 
 By adding other colors such as $g_{P1}-i_{P1}$, $g_{P1}-z_{P1}$, $g_{P1}-y_{P1}$, $r_{P1}-z_{P1}$,
 $r_{P1}-y_{P1}$, $i_{P1}-y_{P1}$ we do not find 
 any improvement in our final results.  
 In summary, we use as meaningful features: 
 the PS1-DR2 PSF-colors (4 features), Kron-colors (4 features),
 PSF-magnitudes (5 features), Kron-magnitudes (5 features) and moments of the radiation intensity (15 features); the AllWISE magnitudes (4 features), magnitudes at different aperture radii (16 features), colors (3 features),
 PS1-DR2/AllWISE mixed colors (45 features), unWISE magnitudes (2 feature),
 PS1-DR2/unWISE mixed colors (5 features).
 We obtain 108 meaningful features in total.
 
We perform random sampling without replacement  to split $90\%$ of the sources in a training set and $10\%$ in a test set. We obtain a training set of 1,126,124 galaxies and  a test set   of 125,125 galaxies. Finally, we normalize the features in the training set according to the formula $X_{st}=(X-\mu)/\sigma$, where $X$ is the input feature, while $\sigma$ and $\mu$ and  are the standard deviation and the mean  of the column feature, respectively.


\begin{table*}
\begin{center}
\begin{tabular}{|c|c|c|c|c|c|c|c|c|}
\hline
 - &$\overline{\Delta z_{norm}}$ &  $\sigma(\Delta z_{norm})$      & $P_{0}$  & $<{\Delta z_{norm}}>$ & $\sigma_{MAD}$  &   $\overline{\Delta z_{norm}}^{\prime}$ & $\sigma(\Delta z_{norm})^{\prime}$  &  $O$    \\
\hline
This work $RF_{cla}$ & $1.0 \times 10^{-3}$   & 0.0225  &  $1.48 \%$ & $2.4 \times 10^{-3}$ & 0.01914 &  $1.8 \times 10^{-3}$ & 0.0252 & 0.34\% \\
This work $RF_{reg}$ & $2.8 \times 10^{-4}$   & 0.0209  &  $1.42 \%$ & $8.2 \times 10^{-4}$ & 0.01764 & $5.3 \times 10^{-4}$  & 0.0235 & $0.22  \%$ \\
\citet{Tarrio2020} & $-2.0 \times 10^{-4}$  & 0.0298 & $4.32 \%$  & -- & -- & -- & -- & --   \\
\citet{Pasquet2019} & --  & -- & -- & $1.0 \times 10^{-4}$ & 0.00912 & -- & -- & --    \\
\citet{Beck2020}  & --  & -- & -- & -- & -- &  $5.0 \times 10^{-4}$ & 0.0322 & $1.89  \%$   \\

\hline
\end{tabular}
\caption{Comparison between the results obtained in this work for both the $RF_{cla}$ and $RF_{reg}$ models with the findings of \citet{Tarrio2020}, \citet{Pasquet2019}  and \citet{Beck2020}. The evaluation metrics are defined in \S \ref{ML2}. \citet{Pasquet2019} achieved a lower $\sigma_{MAD}$ compared to our work. However, our method has the advantage of not relying on SDSS photometry and can be scaled to the entire PS1-DR2 dataset to create a PS1-DR2 photo-$z$ catalog.}
\label{Tab}
\end{center}
\end{table*}

\begin{table*}
\begin{center}
\begin{tabular}{|c|c|c|c|c|c|c|}
\hline
 - &$\overline{\Delta z_{norm}}$ (1)&  $\sigma(\Delta z_{norm})$  (1)    & $P_{0}$ (1)&
 $\overline{\Delta z_{norm}}$ (2)&  $\sigma(\Delta z_{norm})$ (2) &   $P_{0}$ (2)  \\
\hline
This work, $RF_{cla}$ &  $1.1 \times 10^{-3}$  &  0.0238 &  0.014 & $1.5 \times 10^{-3}$ & 0.0253 & 1.31\% \\
This work, $RF_{reg}$ &  $3.4 \times 10^{-4}$  &  0.0220 & 0.014  & $3.8 \times 10^{-4}$ & 0.0234 & 1.42\%  \\

\hline
\end{tabular}
\caption{Comparison between the $RF_{cla}$ and $RF_{reg}$ models trained with just a subsample of meaningful features. (1) models trained with the PS1-DR2/AllWISE features. (2) models trained with the PS1-DR2 features only.  The results obtained with the RF models trained with  the PS1-DR2/AllWISE/unWISE features are presented in  the firsts two lines of Table \ref{Tab}. 
}
\label{Tab1}
\end{center}
\end{table*}

\subsection{Machine learning: a classification vs. regression approach}
\label{ML2}
The supervised ML  goal  consists  of learning a mapping function 
between an input and an output based on example input-output pairs. 
ML models can perform regression and classification tasks. Classification models discriminate objects in two or more classes by ``learning"  a mapping function from a training set and then applying the mapping to  unseen data.  A regression  model  approximates a mapping function  from input variables  to a continuous output variable.
A wide variety of ML models have been constructed, and several have been used in astronomy as well.
In this paper we adopt the random forest (RF)  model
\citep[e.g.,][]{Hastie2010} that is known to be one of the ML models that produces the lowest mean the lowest mean square error (MSE, \citealt{Henghes2021}).

The RF model  is an ensemble ML method for  regression and classification  that works by creating multiple decision trees during training.
The RF algorithm can be used for solving regression ($RF_{reg}$) and classification ($RF_{clas}$) problems \citep[e.g.,][]{Hastie2010}.
The  RF model  depends on multiple hyperparameters: (i) the number of trees in the forest;  (ii) the minimum number of samples needed to split a  node;
(iii) the maximum possible depth of a tree; (iv) the minimum number of samples needed to be a leaf node; (v) a metric  measure of the quality of the split.

ML photometric redshift estimation can be considered a regression problem since given some features (magnitudes, colors and moments), we want to predict a positive real number (the redshift). In a RF model, each tree outputs a redshift value and the final output value is simply the mean value of the trees. This regression problem can be remapped to a classification problem as done by \citet{Pasquet2019}. 
\citet{Pasquet2019} subdivided the  redshift distribution in $K$ equally spaced bins 
 \begin{equation}
 z_{phot}= \sum_{k} z_{k} P(z_{k}),
 \end{equation}
 where $P(z_{k})$ is the probability density function (PDF) of the redshift. The optimal $K$  value  should not be too large to ensure that each bin retains a sufficiently large number of sources, and should not be too small to sample a sufficiently large values of redhifts values.
 We explored several values of $K$ for our subsequent analysis 
 and we find that values of $K\lesssim 15$ lead to a larger  $\Delta z_{norm}$ and $\sigma(\Delta z_{norm})$ (see below for the definition).
 Values of $K \gtrsim 25$ lead to a larger computational time without  an improvement in the results in terms of $\Delta z_{norm}$ and $\sigma(\Delta z_{norm})$. Hereafter, we use $K=20$. 
 
 In this paper we adopt and compare both the RF-regression ($RF_{reg}$) and the RF-classification ($RF_{clas}$) models.
 We use  RF  models with 100 trees. 
 In order to  quantify the  performance of the models we introduce several metrics that have been used in
the photo-$z$ literature that leverages the PS1 and SDSS photometry.
 In \S \ref{main} we use these metrics to test our RF model and to compare our results with other photo-$z$ efforts in literature.

\begin{itemize}
\item One common metric used by \citet{Beck2016} and \citet{Tarrio2020} is the normalized redshift defined as: $\Delta z_{norm}  \equiv  (z_{phot}-z_{spec})/(1+z_{spec})$, where $z_{phot}$ and $z_{spec}$  are the photometric and the spectroscopic redshift, respectively.  Furthermore, after  removing the
outliers defined as $|\Delta z_{norm}| > 3\sigma(\Delta z_{norm})$  (following \citealt{Beck2016} and \citealt{Tarrio2020})
the average bias  is defined as $\overline{\Delta z_{norm}}$.
We also define the outliers rate ($P_0$) as the  fraction of galaxies with  $|\Delta z_{norm}| > 3\sigma(\Delta z_{norm})$.
\item \citet{Pasquet2019} defined the average bias as $<{\Delta z_{norm}}>$ without
removing the outliers and the standard deviation as $\sigma_{MAD}=1.4826 \times MAD$, where MAD (median absolute deviation) is defined as
$\mid \Delta z_{norm}-Median(\Delta z_{norm})\mid$.
\item \citet{Beck2020} defined the outliers  ($O$) as the fraction of galaxies with $|\Delta z_{norm}| > 0.15$. 
These authors computed the average bias ($\overline{\Delta z_{norm}}^{\prime}$)
and the standard deviation ($\sigma(\Delta z_{norm})^{\prime}$)
by removing the outliers ($O$). 

\end{itemize}

The values of these metrics for the photo-$z$ works mentioned in this section are reported in Table \ref{Tab}.

\subsection{Oversampling the training set}
\label{over}
\begin{figure}
\centering
\includegraphics[width=0.5\textwidth]{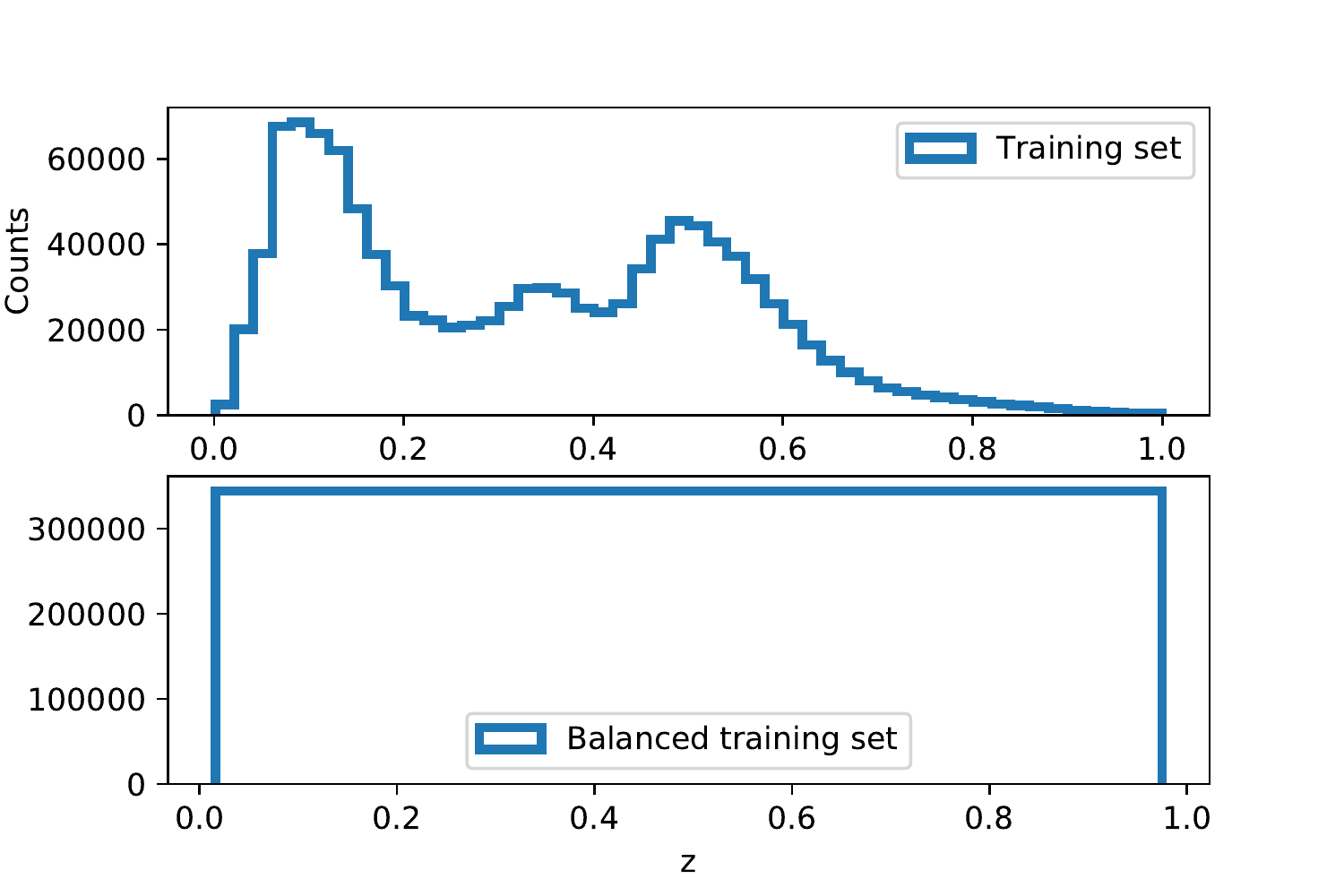}
\caption{
Upper panel: redshift distribution in the unbalanced  training set.
Lower panel: redshift distribution in the oversampled  training set.
Number of sources in the unbalanced training set: 1,126,124.
Number of sources in the oversampled  training set: 3,325,940.
In this work we train an $RF_{reg}$ in the unbalanced training set while we train  a $RF_{clas}$ in the oversampled  training set.
}
\label{fig:00}
\end{figure}
The redshift distribution of the training/testing set is not uniform as displayed in  Figure \ref{fig:00}. Since we subdivided the  redshift distribution in 20 classes, the number of galaxies per class is highly unbalanced. ML classification algorithms  usually perform better on a balanced training set (i.e. same number of elements in each class).  Oversampling algorithms balance the number of sources
in each class by randomly duplicating the sources in the minority classes
until the number of sources in the  minority classes is equal to that in the majority class. The simplest oversampling technique consists of  randomly sampling with replacement the minority classes but this method can result in 
overfitting. To minimize overfitting we employ the  Synthetic Minority Oversampling Technique (SMOTE, \citealt{Chawla2011}).

SMOTE  selects sources that are close in the feature-space, drawing a line between the examples in this space and drawing a new  point along this line.
Specifically, a random source from the minority class is first selected. Then  a number $J$  of the nearest neighbors for that source are found. A randomly selected neighbor (among the J options) is then selected 
and a synthetic example is created at a randomly selected point between the two examples in feature space. In this paper we oversample the minority classes with  Borderline-SMOTE \citep{Han2005}
that is considered an improvement with respect to the standard SMOTE algorithm \citep{Han2005}. 
It is important to mention that we only  oversample  the training set 
and we do not oversample the test set. Figure \ref{fig:00}  shows the  unbalanced redshift distribution of galaxies in the training set (before oversampling, upper panel), and of the oversampled training set (i.e. balanced, lower panel).
The oversampled training set contains 3,325,940 galaxies. In \S\ref{mainqq} we use the oversampled training set to train an RF classification algorithm and the unbalanced training set for the RF regression model.

 \section{Analysis and results}
\label{mainqq}
\begin{figure}
\includegraphics[scale=0.35]{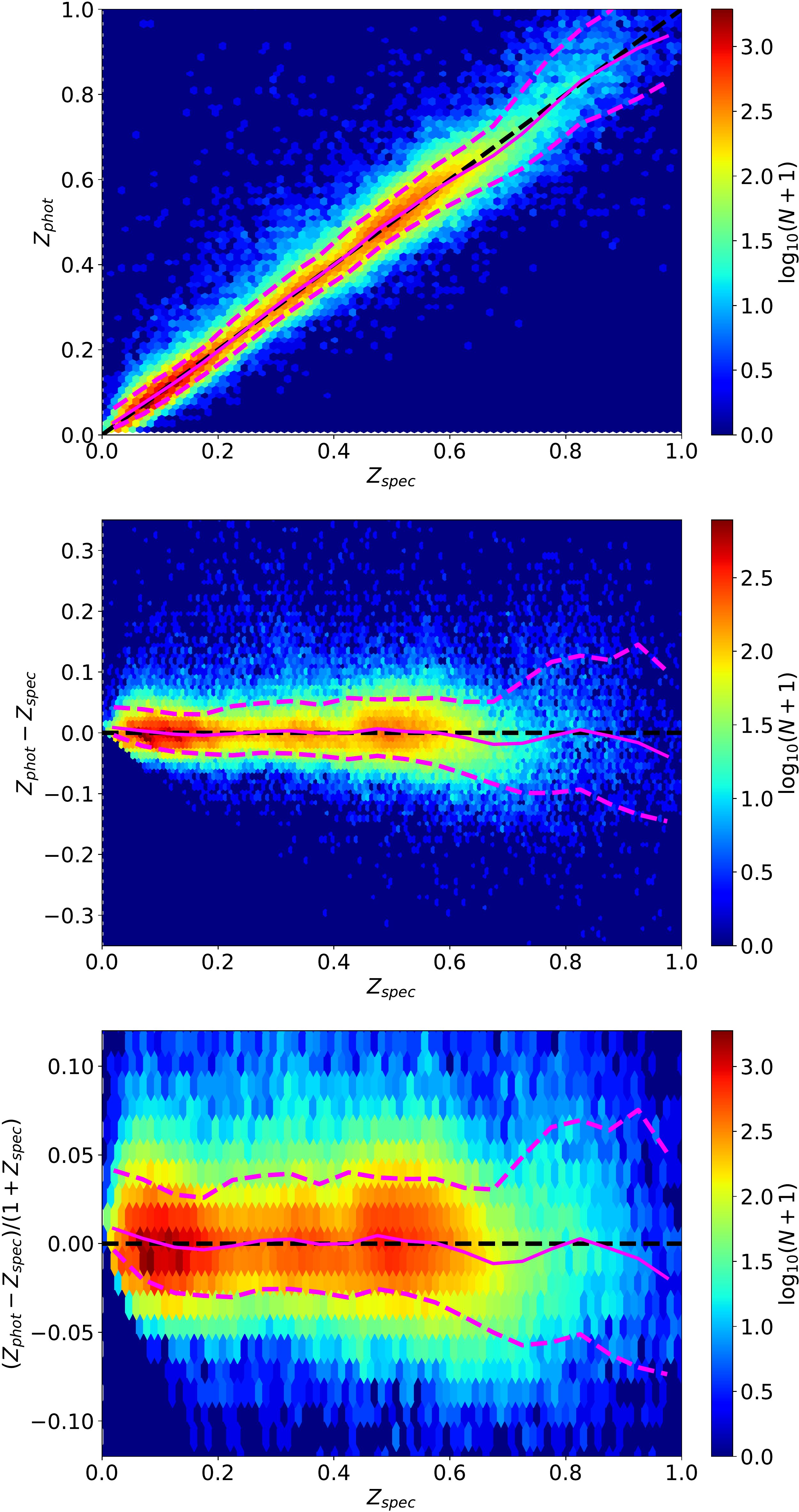}
\caption{ 
Comparison between the photometric redshift $z_{phot}$ estimated in the 
test set with the spectroscopic SDSS redshift $z_{spec}$. 
\emph{Upper Panel}: $z_{phot}$ vs $z_{spec}$.
\emph{Central panel}: $z_{phot}{-}z_{spec}$ vs $z_{spec}$.
\emph{Lower Panel}: $(z_{phot}{-}z_{spec})/(1{+}z_{spec})$ vs $z_{spec}$.
The black dashed line corresponds to  $z_{phot}=z_{spec}$.
Magenta solid line: median value. 
The magenta dashed lines are the tenth and ninetieth percent quantile, respectively. This plot was obtained by applying the $RF_{clas}$ model to the  test set. 
}
\label{fig:1}
\end{figure}

\begin{figure*}
\centering
\includegraphics[width=\textwidth]{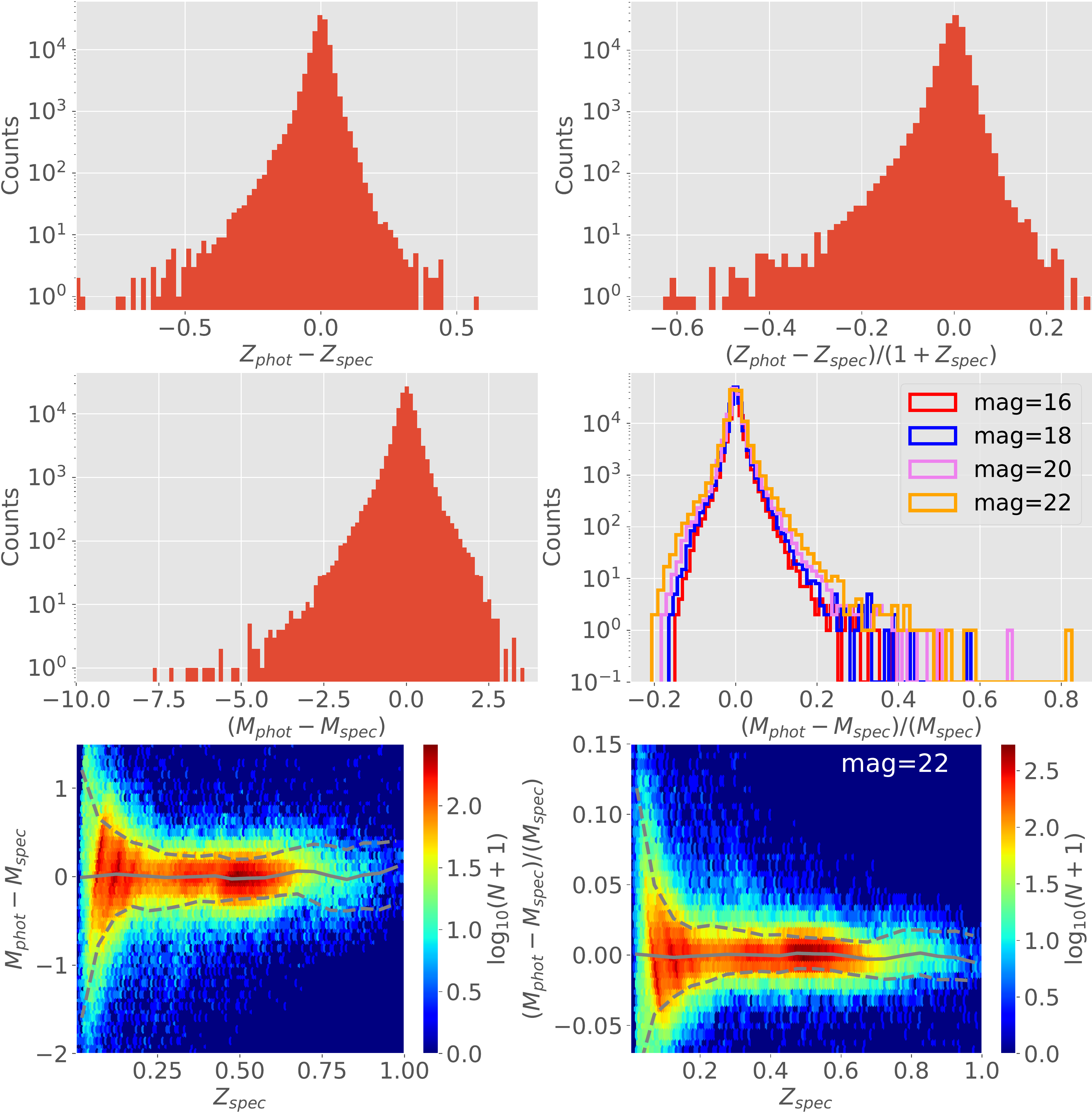}
\caption{
\emph{Upper panels}: distribution of the residuals (left) and  normalized residuals (right) in the test set for the redshift. \emph{Central panels}:  distribution of the residuals  for the absolute magnitudes (left) and  distribution of the normalized residuals (right)  for the absolute magnitudes estimated at different apparent mags. \emph{Lower panels}: (left) $M_{phot}-M_{spec}$ vs. $z_{spec}$ and
(right)  $(M_{phot}-M_{spec})/(M_{spec})$ vs. $z_{spec}$ assuming an apparent mag$=22$. Grey solid line: median value. Dashed grey lines mark the $10^{\rm{th}}$ and $90^{\rm{th}}$ percent quantile, respectively.
This plot was obtained by applying the $RF_{clas}$ model to the  test set.
}
\label{fig:2}
\end{figure*}

\begin{figure}
\includegraphics[width=0.5\textwidth]{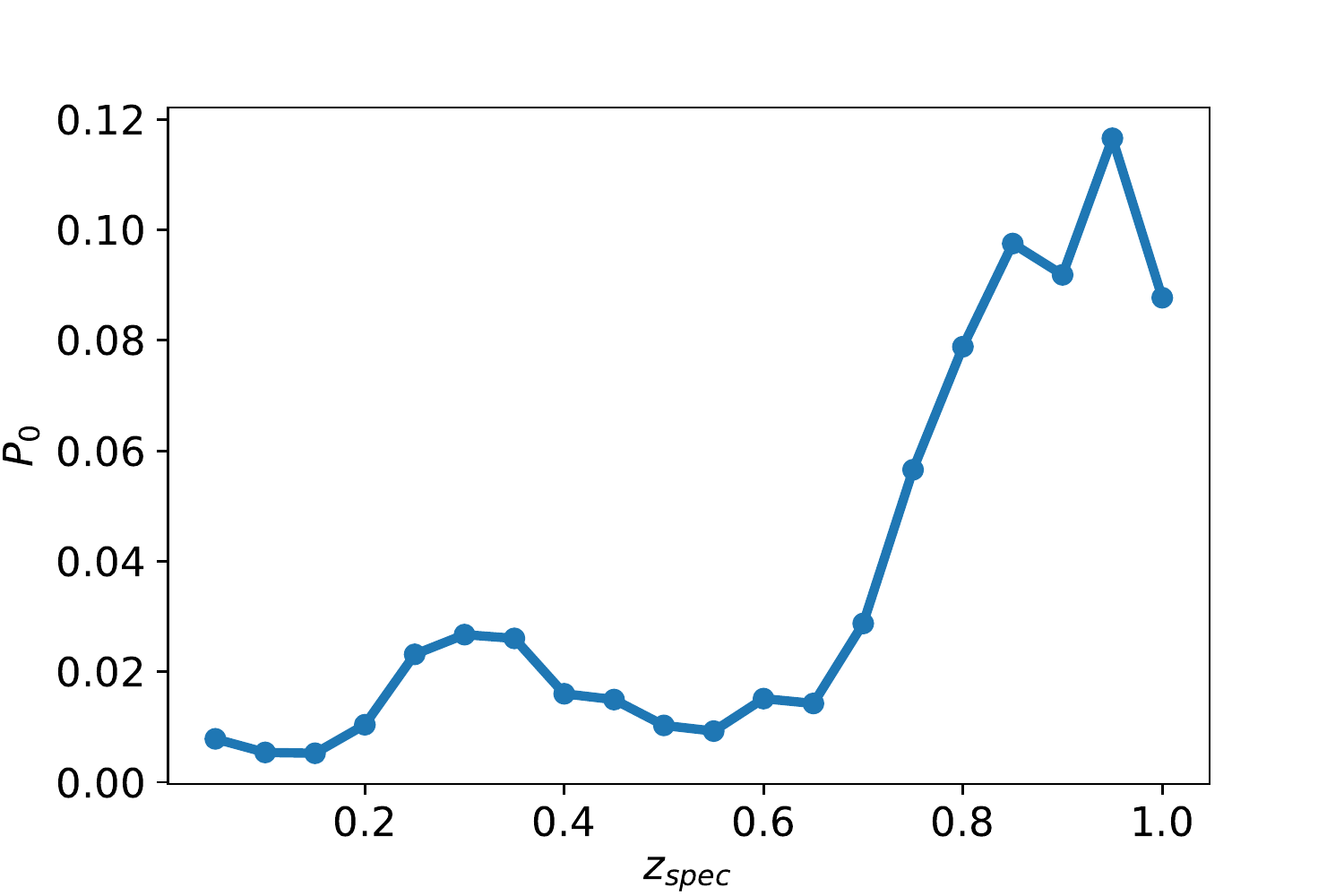}
\caption{Outliers fraction ($P_0$) as a function of 
$z_{spec}$. 
The outliers fraction increases for $z_{spec}>0.7$.
Outliers:  sources with $|\Delta z_{norm}| > 3\sigma(\Delta z_{norm})$.
This plot was obtained by applying the $RF_{clas}$ model to the  test set. 
}
\label{fig:out_fraction}
\end{figure}

\begin{figure*}
\centering
\includegraphics[width=\textwidth]{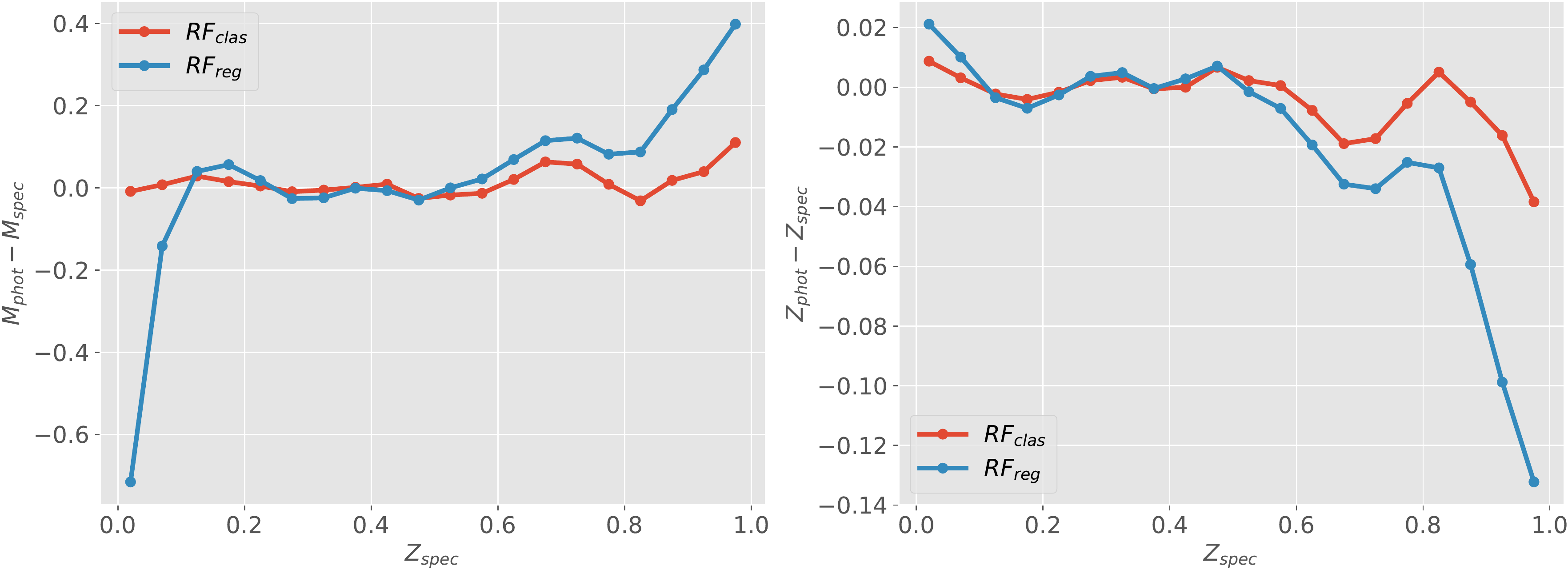}
\caption{Comparison between the $RF_{clas}$ and the $RF_{reg}$ models.
Right panel: median $z_{phot} -z_{spec}$ vs $z_{spec}$ for both $RF_{clas}$
and $RF_{reg}$.
Left panel: median $M_{phot}-M_{spec}$ vs  $z_{spec}$ for both $RF_{clas}$
and $RF_{reg}$.
Median values were calculated by subdividing $z_{spec}$ in 20 equally spaced bins
and we then calculated the median $z_{phot} -z_{spec}$ and $M_{phot}-M_{spec}$ for each of those bins.
The figure shows that $RF_{clas}$ is better performing for $z<0.1$ and $z>0.6$ in terms of medians.
}
\label{fig:conpa}
\end{figure*}

\begin{figure*}
\centering
\includegraphics[width=\textwidth]{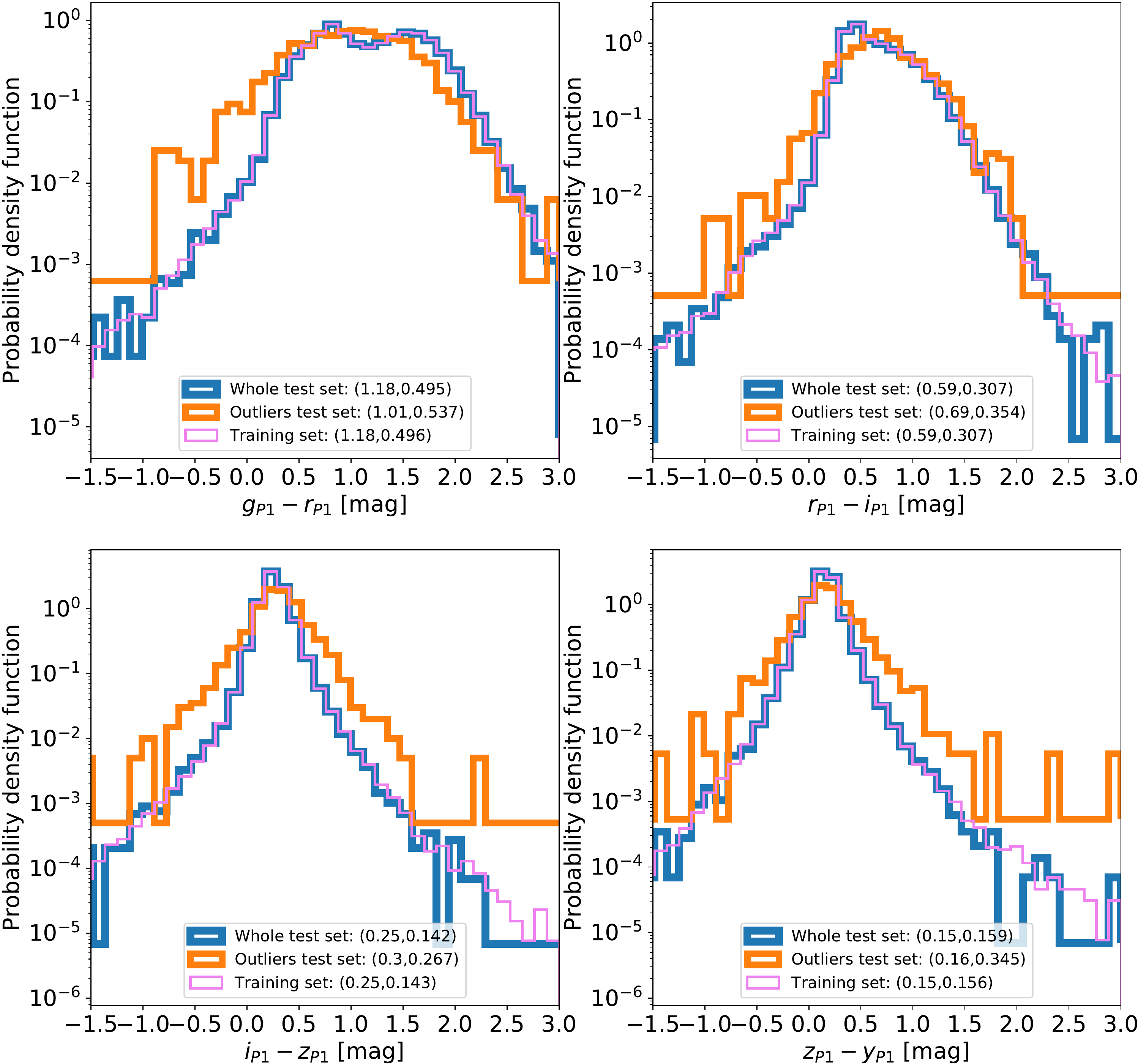}
\caption{
Normalized distribution of $g_{P1}-r_{P1}$, $r_{P1}-i_{P1}$, $i_{P1}-z_{P1}$ and $z_{P1}-y_{P1}$ colors for the galaxies in the test and training set. Blue: Entire  test set. Orange: distribution of colors of  outliers in the  test set, i.e., sources with $|\Delta z_{norm}| > 3\sigma(\Delta z_{norm})$.
Violet: training set.
The numbers within the parenthesis in the legend are the median and the standard deviation of the data, respectively. The standard deviation of the outliers distribution is larger than that of one of the entire test set,    indicating that sources that are outliers are more likely to span a wider range of colors than a typical source in the test set. 
This plot was obtained by applying the $RF_{clas}$ model to the  test set.
}
\label{fig:out}
\end{figure*}

In this section we train and test the $RF_{clas}$ and the $RF_{reg}$ models (\S\ref{main}), and compare their performances. Since the  $RF_{clas}$ model performs better in the local universe $z\le 0.1$, which is our primary interest,  we focus our detailed analysis  of the confidence intervals for the photo-$z$ estimation  (\S\ref{main}) and of the outliers (\S\ref{out}) on the $RF_{clas}$  model. We compare our results with other works in the literature in \S\ref{lit}.
We display in Figures \ref{fig:1}, \ref{fig:2}, \ref{fig:out_fraction} and \ref{fig:outqq} the results relative to the  $RF_{clas}$ algorithm,
and in Figure \ref{fig:conpa} we show a comparison between the  $RF_{clas}$ and  $RF_{reg}$.

\subsection{A photo-z ML model}
\label{main}

We train the $RF_{reg}$ and the $RF_{clas}$  models in the training set and in the oversampled training set, respectively. We test  both  $RF_{reg}$ and  $RF_{clas}$ in the test set. %
Our results for the  $RF_{clas}$ model are:
$\overline{\Delta z_{norm}}=1.0 \times 10^{-3}$,
$\sigma(\Delta z_{norm})=0.0225$
and $P_{0}=0.0148$ in the test set (Table \ref{Tab}).
  The $RF_{reg}$ model yields:
 $\overline{\Delta z_{norm}}=2.8 \times 10^{-4}$,
$\sigma(\Delta z_{norm})=0.0209$
and $P_{0}=0.0142$ in the test set (Table \ref{Tab}).
 The  $RF_{reg}$  globally performs  better than the  $RF_{clas}$ model. However,  as we will show below, the $RF_{clas}$ model performs better for $z<0.1$ that is paramount for
transient classification in the local Universe. 
For this reason we focus the subsequent analysis on the $RF_{clas}$ model. Since the training/testing sets are built through random sampling, we repeat five times  the same procedure described above
to test for the stability of the model.
We do not find any  differences above $1\%$ in $\sigma(\Delta z_{norm})$  in  the different runs.

Figure \ref{fig:1} shows the  comparison  of the photometric redshift $z_{phot}$ inferred with the $RF_{clas}$ model,  with the true spectroscopic redshift $z_{spec}$ in the test set. 
The left panel of Figure \ref{fig:1} shows   $z_{phot}$ as a function of $z_{spec}$ while the middle and right panels show $z_{phot} -z_{spec}$ and 
$\Delta z_{norm}$ as a function of $z_{spec}$, respectively. 
Figure \ref{fig:1} reveals that the $z_{phot}$ forecasting are overall compatible  with $z_{spec}$. The most difficult range of redshift values to predict are  those at  $z\gtrsim0.7$ ( right panel of figure \ref{fig:1}).

The photometric prediction error ($z_{phot} -z_{spec}$) will induce an absolute magnitude prediction error ($M_{phot}-M_{spec}$), where 
$M_{phot}$ and $M_{spec}$ are the absolute magnitude estimated at  $z_{phot}$ and  $z_{spec}$, respectively. By definition, $M_{phot}-M_{spec}=5 \log_{10}(d_{spec}/d_{phot})$, where  $d_{spec}$ and $d_{phot}$
 are the luminosity  distances\footnote{We estimate the distances by using the FlatLambdaCDM class from astropy based on \citet{Planck2016}} to a galaxy estimated at $z_{spec}$ and  $z_{phot}$, respectively.
 In the central left panel of Figure \ref{fig:2} we display the distribution of  $M_{phot}-M_{spec}$ revealing
 that  97\% of the galaxies have  $|M_{phot}-M_{spec}|\le 1$ mag
 and 73\% of the galaxies have $|M_{phot}-M_{spec}|\le 0.3$ mag. We also find that 87\% of the galaxies with $z_{spec}<0.1$ have $|M_{phot}-M_{spec}|\le 1$ mag 
 and 40\%  have $|M_{phot}-M_{spec}|\le 0.3$ mag.
 
 In the lower left panel of Figure \ref{fig:2} we show the distribution of $M_{phot}-M_{spec}$ as a function of $z_{spec}$
 revealing that the median absolute value of the error is 
 approximately 0 for all bins of $z_{spec}$.
 It is also important to explore the distribution of the normalized absolute magnitude error defined as
 $(M_{phot}-M_{spec})/M_{spec}= \log_{10}(d_{spec}/d_{phot})/(1+0.2 \,mag-\log_{10}(d_{spec}/(pc))$, where $mag$ is the apparent magnitude of the galaxy. In the central right panel  of figure \ref{fig:2} we show the distribution of $(M_{phot}-M_{spec})/M_{spec}$ for different $mag$ values revealing that 99\% of the galaxies have a $(M_{phot}-M_{spec})/M_{spec}$ absolute value less than 0.1 for an apparent mag value of 22. 
 Furthermore, in the lower panel of Figure \ref{fig:2} we display the $(M_{phot}-M_{spec})/M_{spec}$ distribution as a function of $z_{spec}$
 revealing that  for $z_{spec}<0.1$ we obtain the largest relative  error value.
 In the upper panels of Figure \ref{fig:2} we also display for reference the distribution of $z_{phot} -z_{spec}$ and 
$\Delta z_{norm}= (z_{phot}-z_{spec})/(1+z_{spec})$, respectively.

In the right panel of Figure \ref{fig:conpa} we show the median value of $z_{phot} -z_{spec}$ as a function  of $z_{spec}$ for both the $RF_{reg}$ and the $RF_{clas}$  models in the test set. The $RF_{clas}$ model is  better performing at $z_{spec}<0.1$ and  $z_{spec}>0.6$ (in terms of median values) while it provides comparable results for $0.1<z_{spec}<0.6$.  A similar result holds when we compare the median value of $M_{phot}-M_{spec}$ as a function of $M_{spec}$ (see left panel of Figure \ref{fig:conpa}) for the $RF_{clas}$ and the $RF_{reg}$ models.
The $RF_{clas}$ model is in median better performing than the $RF_{reg}$ at
$z_{spec}<0.1$ and  $z_{spec}>0.6$ because the $RF_{clas}$ was trained on an oversampled training set whereas the $RF_{reg}$ in a training set with very few sources at very low and very high z (see figure \ref{fig:00}).
If we train the $RF_{clas}$ in the standard (not oversampled) training set we obtain a result that is approximately identical to the one of the $RF_{reg}$ model. Therefore,  the differences between  $RF_{clas}$ and  $RF_{reg}$ are simply due to a different training set.

Next we address the topic of the importance of the features in the models. 
We measure  feature importance by considering a $RF_{reg}$ and a $RF_{clas}$ model with  the PS1-DR2 features only and one model  with the PS1-DR2/AllWISE features. We evaluate the goodness of the model  by estimating $\overline{\Delta z_{norm}}$,
$\sigma(\Delta z_{norm})$ and $P_{0}$ in the test set for the $RF_{reg}$ and for the $RF_{clas}$ models, respectively. 
Quantitative results are summarized in Table \ref{Tab1}.
This experiment shows that by adding the AllWISE and unWISE photometry 
we improve the results for both the $RF_{reg}$ and  $RF_{clas}$, respectively.

We conclude this section with a discussion of the confidence intervals for the photometric redshift estimation with the $RF_{cla}$ model.
We  use the RF trees to build an empirical  probability density function of the photometric redshift  for  each galaxy.
We define the  $80\%$ confidence interval  as
($z_{phot90},z_{phot10}$), where $z_{phot90}$ and $z_{phot10}$ are $90\%$ and the $10\%$ quantile, respectively.
We find that $\approx 80\%$  of the spectroscopic redshift of the  galaxies in the test sets are within the ($z_{phot90},z_{phot10}$) confidence interval, for both  
$RF_{cla}$ and $RF_{reg}$,
indicating that this method yields realistic estimates of the true $80\%$ confidence interval. In a future work we will release a catalog with the photometric redshfit estimates and with the confidence intervals for each of the sources in the PS1-DR2 catalog.

\subsection{Outliers analysis}
\label{out}

In \S\ref{ML} we defined the outliers as the galaxies with $|\Delta z_{norm}| > 3\sigma(\Delta z_{norm})$ following \citet{Tarrio2020}. 
In Figure \ref{fig:out_fraction} we display the fraction 
of outliers as a function of $z_{spec}$ for the sources in the test set. The outliers fraction is larger for $z_{spec}>0.7$.
\begin{figure*}
\centering
\includegraphics[width=\textwidth]{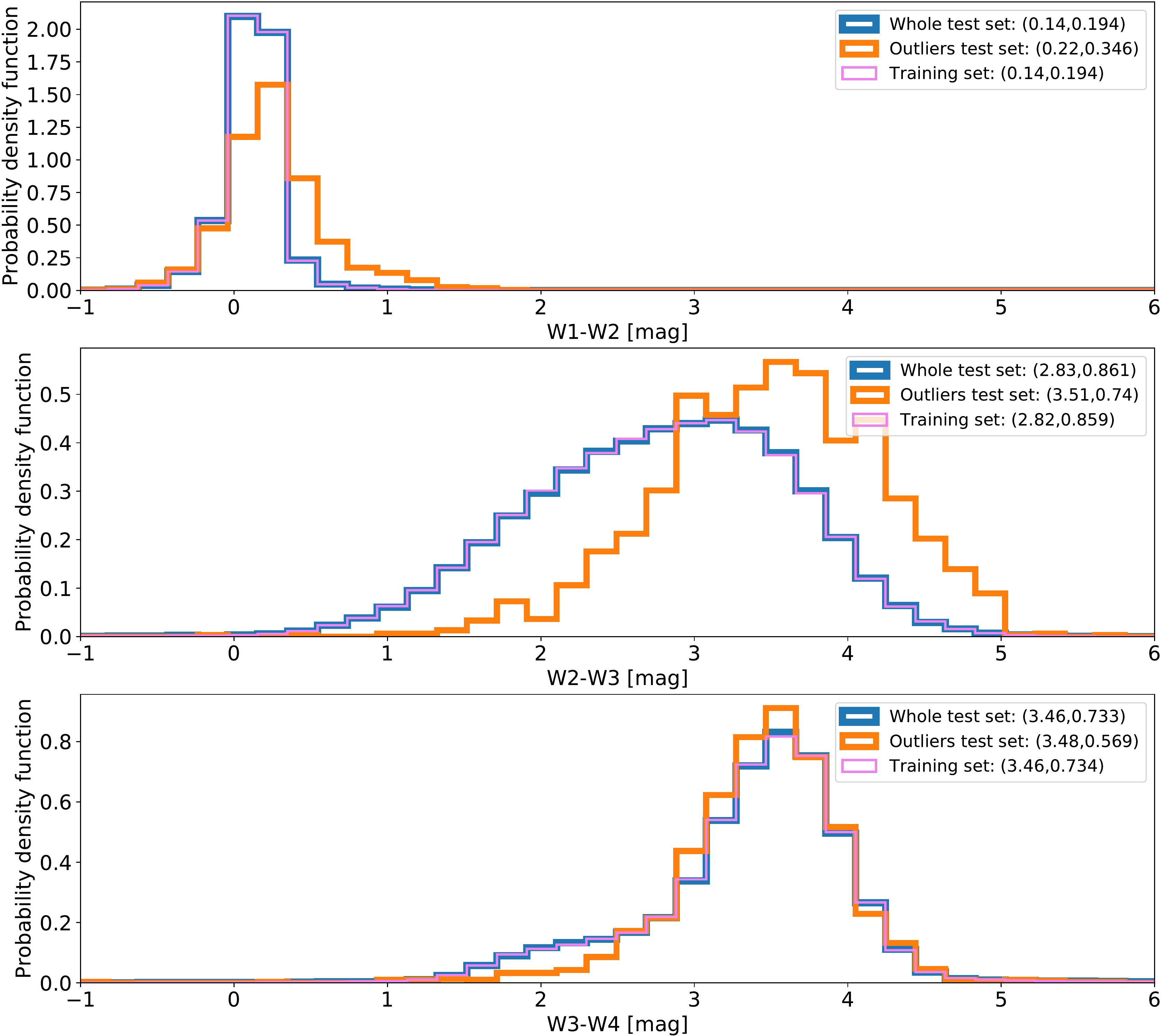}
\caption{Normalized distribution of  W1-W2, W2-W3 and W3-W4 colors for the galaxies in the test and training  set. Blue: entire  test set. Orange:  outliers in the  test set namely sources with $|\Delta z_{norm}| > 3\sigma(\Delta z_{norm})$.  
Violet: training set.
The numbers within the parenthesis in the legend are the median and the standard deviation of the data, respectively. This plot was obtained by applying the $RF_{clas}$ model to the  test set.
}
\label{fig:outqq}
\end{figure*} 

It is meaningful to understand the properties of those outliers when compared to the entire population. In Figure \ref{fig:out} we show the normalized distribution of $g_{P1}-r_{P1}$, $r_{P1}-i_{P1}$, $i_{P1}-z_{P1}$ and $z_{P1}-y_{P1}$ in the test set for the entire sample and for the outliers, respectively.
Figure \ref{fig:out} reveals that the distributions of colors of the outliers  broadly overlap with, but is not identical to the color distributions of the entire test set. We perform a two-tail Kolmogorov-Smirnov test 
to quantify the distance between the outliers and the whole sample. We find that the probability that the two populations share the same parent population is $p_{value}<10^{-10}$ for each of the four colors suggesting that the two populations are different.

The main difference between the entire distribution and the outliers distribution is in terms of the sample standard deviation. The 
outliers appear more scattered than the whole distribution for each of the four colors.
The ouliers and the whole distributions appear not particularly different in terms of median values. In Figure \ref{fig:out} we report the median values and the standard deviations of the distributions.

We repeat the same exercise with IR colors.
Figure \ref{fig:outqq} displays the distribution of the AllWISE colors  of the outliers and of the whole population. Figure \ref{fig:outqq} reveals that the outliers distribution is different from the whole distribution. A two-tail Kolmogorov-Smirnov test leads to a $p_{value}<10^{-6}$ for each of the three colors, suggesting that the outliers and the whole distribution tend to have differen IR colors.
The figure shows that the outliers are  slightly redder  than the whole population  for the W1-W2 color, significantly redder for the  W2-W3 color, while having comparable median values for the W3-W4 color. 

\begin{figure}
\includegraphics[width=0.5\textwidth]{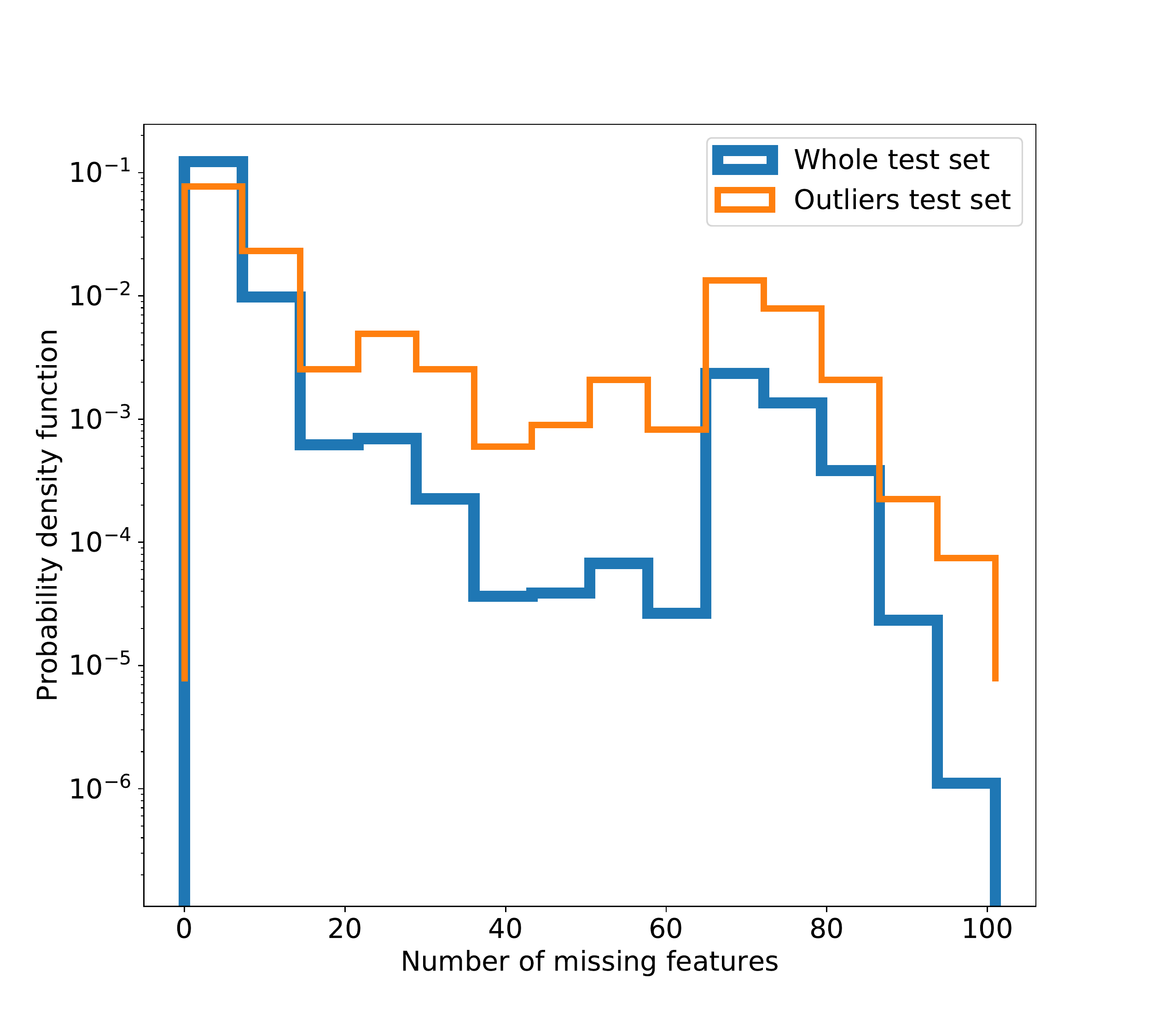}
\caption{ Normalized distribution of the number of missing features for the galaxies  in the test set.
Blue thick line: histogram for the entire  population. Orange thick line: histogram of the outliers, namely sources with $|\Delta z_{norm}| > 3\sigma(\Delta z_{norm})$.
Mean value for Orange: 17.6. Mean value for Blue: 3.2.
Outliers galaxies have on average a larger number of missing features, as expected.
}
\label{fig:out2}
\end{figure}

 Our dataset contains  galaxies with missing features. The presence of missing feature often leads to poorer performances of a ML model. Thus, we compare the distribution of missing features in the entire population with the outlier sample in Figure \ref{fig:out2}. As expected, Figure \ref{fig:out2} reveals that on average the outliers have  a larger number of missing features than the whole population, 3.2 and 17.6 for whole population and outlier sample, respectively, which leads to a poorer performance of the ML algorithms for the outlier sample.
In conclusion the outliers have on average (1) a higher number of missing features  and (2) colors that are not well represented in the training set.

\subsection{Comparison with  the literature}
\label{lit}
Machine learning is playing an important role in photo-$z$ estimation in astronomy.
A detailed  comparison  between all the results of this work and other papers from the literature is not always possible as each study uses different data catalogs, features and redshift distributions for training and testing.  Here we offer a quantitative comparison of our results and results from the literature  that use the PS1 or the SDSS datasets using the metrics defined in \S\ref{ML2} and summarized in Table \ref{Tab}. The most relevant  metric to evaluate the performance  of a photo-$z$ model  is the standard deviation that can be expressed as  $\sigma(\Delta z_{norm})$, $\sigma(\Delta z_{norm})^{\prime}$ and 
$\sigma_{MAD}$.

The study that is more directly comparable with ours is  \citet{Tarrio2020}, where the authors used  SDSS-DR16 and PS1-DR2. \citet{Tarrio2020} 
used SDSS-DR16 spectroscopic labels as training labels and 
PS1-DR2 colors as as features. \citet{Tarrio2020} adapted a linear regression method developed by \citet{Beck2016} for the photo-$z$ estimation
and obtained    $\overline{\Delta z_{norm}}=-2.01 \times 10^{-4}$, a standard deviation $\sigma(\Delta z_{norm})=0.0298$
and an outlier rate of  $P_0=4.32 \%$ in the test set.
 We note that while \citet{Tarrio2020} tested their model with  galaxies  without missing values   (no missing PS1-DR2 colors),  we include the galaxies with missing features when we calculated our metrics.  In spite of this, we obtain a smaller $\sigma(\Delta z_{norm})$,   a smaller $P_{0}$ and a slightly larger $\overline{\Delta z_{norm}}$
 than \citet{Tarrio2020} with both the $RF_{reg}$ and $RF_{clas}$ models (see table \ref{Tab}).

\citet{Beck2020}   leveraged the PS1-DR1 features and SDSS spectroscopic redshift for photo-$z$ estimation. \citet{Beck2020} used a deep neural network for  photo-$z$ estimation obtaining $\overline{\Delta z_{norm}}^{\prime}=5 \times 10^{-4}$, $\sigma(\Delta z_{norm})^{\prime}=0.0322$ and $O= 1.89 \%$.
 We obtain a smaller $\sigma(\Delta z_{norm})^{\prime}$, $O$ and  $\overline{\Delta z_{norm}}$ than \citet{Beck2020} with both  $RF_{reg}$ and $RF_{clas}$ models.

\citet{Pasquet2019} adopted  a convolutional neural network approach on the SDSS-DR16 dataset for photo-$z$ estimation using the SDSS-DR16 spectroscopic redshift as labels and the 
DSS-DR16 images as features. \citet{Pasquet2019} obtained  $<{\Delta z_{norm}}>=0.0001$ and $\sigma_{MAD}=0.00912$.
Even though \citet{Pasquet2019} achieved a lower $\sigma_{MAD}$ compared to our work, our method has the benefit to be directly applicable to the 
entire PS1-DR2 dataset.

\section{Summary and Conclusions}
\label{conclusions}
We present two machine-learning models to compute photometric redshifts (photo-$z$)
for galaxies.  We use data from  PS1-DR2, AllWISE/unWISE and SDSS-DR16.
Our method relies  on a random-forest  regression ($RF_{reg}$) and on a random-forest classification ($RF_{clas}$)
algorithm leveraging the PS1-DR2/AllWISE/unWISE features (colors, magnitudes and moments of the radiation intensity)
and the SDSS-DR16 labels (spectroscopic redshift).
The $RF_{clas}$ was trained using an oversampled training set to 
equally weigh the underrepresented portion of the redshift distribution.
We 
obtained
$\overline{\Delta z_{norm}}=1.0 \times 10^{-3}$,  $\sigma(\Delta z_{norm})=0.0225$ and $P_0=1.48 \%$ for the $RF_{clas}$ model, 
and $\overline{\Delta z_{norm}}=2.8 \times 10^{-4}$,  $\sigma(\Delta z_{norm})=0.0209$ and $P_0=1.42 \%$ for the $RF_{reg}$ model, respectively.

We  analyze the photo-$z$ estimation as a function of the spectroscopic redshift finding 
that the largest photo-$z$ error ($z_{phot} -z_{spec}$) is for low redshifts ($z\lesssim 0.1$) and for high redshifts ($z\gtrsim0.6$) for the $RF_{reg}$ model.
The $RF_{clas}$ model  performs better than the $RF_{reg}$ model in the local Universe ($z\lesssim0.1$) and at high redshift ($z\gtrsim0.6$). This difference is mostly a consequence of the fact that the $RF_{clas}$, differently from the $RF_{reg}$ model, was trained in a oversampled training set. 

Furthermore, we explore how the photo-$z$ error affects the absolute magnitude photometric error, defined as $M_{phot}-M_{spec}$. We find that $97 \%$ of the galaxies have $\mid M_{phot}-M_{spec} \mid \le 1$ mag suggesting that our photo-$z$ can be used for course transient classification. We also find that 87\% of the galaxies with $z_{spec}<0.1$ have $|M_{phot}-M_{spec}|\le 1$ mag.
 In a follow-up paper we will release a catalog with the photo-$z$ estimation of the entire PS1-DR2 dataset.

\bigskip
\bigskip\bigskip\bigskip
\section*{Acknowledgments}
This work is supported by the Heising-Simons Foundation under grant \#2018-0911 (PI: Margutti).
R.M. is grateful to KITP for hospitality during the completion of this paper. This research was supported in part by the National Science Foundation under Grant No. NSF PHY-1748958.
R.M. acknowledges support by the National Science Foundation under Awards No. AST-1909796 and AST-1944985. Raffaella Margutti is a CIFAR Azrieli Global Scholar in the Gravity \& the Extreme Universe Program, 2019. 
A.A.M.~is funded by the Large Synoptic Survey Telescope Corporation, the
Brinson Foundation, and the Moore Foundation in support of the LSSTC Data
Science Fellowship Program; he also receives support as a CIERA Fellow by the
CIERA Postdoctoral Fellowship Program (Center for Interdisciplinary
Exploration and Research in Astrophysics, Northwestern University).
The Pan-STARRS1 Surveys (PS1) and the PS1 public science archive have been made possible through contributions by the Institute for Astronomy, the University of Hawaii, the Pan-STARRS Project Office, the Max-Planck Society and its participating institutes, the Max Planck Institute for Astronomy, Heidelberg and the Max Planck Institute for Extraterrestrial Physics, Garching, The Johns Hopkins University, Durham University, the University of Edinburgh, the Queen's University Belfast, the Harvard-Smithsonian Center for Astrophysics, the Las Cumbres Observatory Global Telescope Network Incorporated, the National Central University of Taiwan, the Space Telescope Science Institute, the National Aeronautics and Space Administration under Grant No. NNX08AR22G issued through the Planetary Science Division of the NASA Science Mission Directorate, the National Science Foundation Grant No. AST-1238877, the University of Maryland, Eotvos Lorand University (ELTE), the Los Alamos National Laboratory, and the Gordon and Betty Moore Foundation.

\bibliographystyle{aasjournal}
\bibliography{master_sne}


\end{document}